\documentstyle[aps,prl]{revtex}
\begin{document}
\author{E. A. Jagla and C. A. Balseiro}
\address{{\it Comisi\'on Nacional de Energ\'\i a At\'omica}\\
{\it Centro At\'omico Bariloche and Instituto Balseiro}\\
{\it 8400, S. C. de Bariloche, Argentina}}
\title{Percolation transition of the vortex lattice and {\it c}-axis
resistivity in
high-temperature superconductors}
\maketitle

\begin{abstract}
We use the three-dimensional Josephson junction array system as a model for
studying the temperature dependence of the {\it c}-axis resistivity of high
temperature superconductors, in the presence of an external magnetic field $%
H $ applied in the {\it c}-direction. We show that the temperature at which
the dissipation becomes different from zero corresponds to a percolation
transition of the vortex lattice. In addition, the qualitative features of
the resistivity vs. temperature curves close to the transition are obtained
starting from the geometrical configurations of the vortices. The results
apply to the cases $H\neq 0$ and $H=0$.
\end{abstract}


\narrowtext

Strong thermal fluctuations and anisotropy make the physics of the vortex
lattice in high-Tc materials be much more rich and complicated than
predictions of mean field theories\cite{blatter}. This shows up, in
particular, in the complicated structure of the field-temperature ($H$-$T$)
phase diagram of the high-Tc's. It seems to be clear that there is a line in
the $H$-$T$ phase diagram that separates a low-temperature phase (known as
the vortex glass phase\cite{vg}) where vortex lines are frozen in space, and
a high-temperature phase in which the vortex lines move through the material
due to thermal activation. The passage to the normal state when increasing
temperature is likely to be a crossover, instead of a well-defined
transition. The curve that separates the low- and high-temperature phases is
named irreversibility line (IL). The $V$-$I$ characteristics when an
external current is applied perpendicularly to the magnetic field is
different above and below the IL. Below IL the $V$-$I$ curves are well
fitted by $V\sim \exp \left[ (-I_c/I)^\mu \right] $ with $\mu $ and $I_c$
being two parameters\cite{vg,vg2}, and in particular, the resistivity $\rho $
of the system -which is defined as $\rho =\lim _{I\rightarrow 0}(V/I)$- is
strictly zero. Above IL the behavior is ohmic, i.e., $V\sim I$ \cite{actterm}%
{}.

When the current is applied parallel to the field, the mean force exerted on
the vortices is zero. However there are local forces -due to misalignment of
the local magnetic field- that may give rise to dissipation. The most
important mechanism for dissipation in this configuration at intermediate
temperatures is the thermal activation of vortex loops, which gives a
voltage $V_c\sim \exp \left[ -I_c/I\right] $\cite{blatter}, implying zero
resistivity. In this work we show that when temperature is increased there
is a phase transition at a temperature $T_p$ that reflects a thermodynamic
property of the vortex system and is signed by the occurrence of a non-zero
resistivity. In fact, the $I-V$ characteristic for YBaCuO, when current and
magnetic field are parallel to the {\it c}-axis show the following behavior
\cite{ejecexp,ejecexp2}: for small currents and high temperature the
response is ohmic, the range of currents that gives a linear response is
reduced as the temperature decreases, and at a well-defined temperature $T_p$
the linear behavior disappears. Moreover, the $I-V$ curves can be scaled on
two universal curves, corresponding to $T>T_p$ and $T<T_p$ respectively \cite
{privado}. This behavior -that is similar to what occurs when the current is
applied in the {\it ab}-plane- supports the idea of a thermodynamic
transition that we identify with a percolation transition of the vortex
system. Experimentally, it is observed\cite{ejecexp} that the dissipation in
the {\it c}-axis appears at different temperatures than in the {\it ab}%
-plane. This implies that the `irreversibility line' for a current parallel
to the field is different than the corresponding to the {\it ab}-plane.

Here we explore the following idea. At zero temperature, the vortices are
straight lines, and the net force on each of them when a small current in
the {\it c}-axis is present is zero. At low temperature, vortex lines start
to wander and vortex loops are created due to thermal activation. However,
if temperature is not too high, vortex loops and vortex lines are still
isolated from each other and dissipation in the linear regime is zero
-except for surface effects (see below). When increasing temperature, vortex
lines and thermally generated vortex loops start to touch each other and for
temperatures greater than a critical value $T_p$, there will be a vortex
path crossing the sample along the {\it ab}-plane. The net force exerted by
the current on this path is different from zero, and a finite dissipation
will be observed. In this way we qualitatively see that the existence of
paths perpendicular to the current in the sample -i.e., the transversal
percolation of the vortex lattice- is crucial for the dissipation in the
{\it c}-axis\cite{perc}.

The model used to test this idea is the three dimensional Josephson junction
array on a discrete lattice, that has been described in detail elsewhere\cite
{djb,modelo}. The dynamics of the model is contained in the evolution of the
phases $\varphi ^i(t)$, which are defined on the nodes of a cubic lattice
and represent the phase of the order parameter. Between nearest neighbors
nodes there are Josephson junctions characterized by a critical current $I_0$
and a normal resistance $R_0$. The equations describing the model are

\begin{equation}
\label{corrt}j^{ii^{\prime }}=I_0\sin \left( \varphi ^i-\varphi ^{i^{\prime
}}-A^{ii^{\prime }}\right) +\frac 1{R_0}\frac{\partial (\varphi ^i-\varphi
^{i^{\prime }})}{\partial t}+\eta ^{ii^{\prime }}(t)\text{ }
\end{equation}

\begin{equation}
\label{kirch}\sum_{\{i^{\prime }\}}j^{ii^{\prime }}=j_{ext}^i.
\end{equation}

Eq. \ref{corrt} gives the current $j^{ii^{\prime }}$ between nearest
neighbors nodes $i$ and $i^{\prime }$ with phases $\varphi ^i$ and $\varphi
^{i^{\prime }}$. Here $A^{ii^{\prime }}$ is the vector potential of the
external magnetic field, and $\eta ^{ii^{\prime }}(t)$ is an uncorrelated
gaussian noise which incorporates the effect of temperature. Eq. \ref{kirch}
assures the current conservation on each node, and $j_{ext}^i$ is the
external current applied at node $i$.

The model allows for the existence of vortices, which consist in
singularities of the phases $\varphi (t)$ around a given closed path.
Self-inductance and disorder effects are not considered and the system is
taken isotropic for simplicity -i.e., $I_0$ and $R_0$ are taken constant
throughout the lattice.

We numerically integrate Eqs. \ref{corrt},\ref{kirch} in time. Voltages at
different points of the sample are calculated as the temporal mean value of
the time derivative of the phases $\varphi $. The resistivity of the sample
in a given direction is calculated by injecting a small external current
(typically around $\sim 1/20$ of the critical current of the junctions) by
one of the faces of the sample and withdrawing it from the opposite face.
The small value of the external current is chosen in order to be in the
linear regime, in which the voltage drop is proportional to the applied
current.

The boundary conditions (BC) are taken open in the {\it ab}-plane. However,
if open BC in the {\it c}-axis are used, there will be a finite force on an
isolated vortex at finite temperature if the top and bottom ends of the
vortex are not aligned. The dissipation -that is non-zero even in the linear
regime- caused by this net force turns out to be independent of the
thickness of the sample\cite{modelo}, and in this sense, it is only a
surface effect. In order to eliminate this spurious surface effect it is
crucial to use BC for the {\it c}-direction that assure that each vortex
line leaving the sample at a given point of the bottom plane re-enters at
the same point of the top plane. Strict periodic BC on the phases $\varphi $
have this property, however we would obtain that the voltage difference
between top and bottom planes is identically zero. We use, instead, open BC
for the mean value of the phases in the top ($\bar {\varphi _T}$)and bottom (%
$\bar {\varphi _B}$) planes, and periodic BC for all the phase differences $%
\varphi _T^i-\varphi _T^j$ and $\varphi _B^i-\varphi _B^j$. This guarantees
the periodicity of the vortex configurations and permits the calculation of
the {\it c}-axis resistivity.

We have to define a criterium for percolation: in our model there is a
typical length which is the lattice parameter $a$. Distances smaller than $a$
cannot be resolved. Flux conservation implies that every flux line going
into a unit cell of our lattice also goes out of the cell. When two vortices
go into the same elemental cell we cannot tell which one of the two outgoing
vortices correspond to each one of the ingoing vortices. We interpret this
situation as the meeting of two vortex lines. In a real material this
corresponds to two vortex lines being at a distance lower than the core size
of the vortex. At high enough temperatures the vortex structure may
percolate perpendicularly to the applied field: starting from one side of
the sample we can follow a vortex line and arrive to the opposite side of
the sample. Due to the finite size of the systems used, and to the dynamical
evolution, percolation is not expected to occur at every time, but only at a
given fraction of the total time, which depends on temperature. We evaluate
the probability that there exists a vortex line crossing the system from one
side to the opposite as a function of temperature. Because a sharp
percolation transition can only be seen in the thermodynamic limit, we do
scaling with the size of the system.

In Fig.\ref{volperc}(a) we show the resistivity of a cubic ($L_{ab}\times
L_{ab}\times L_c$, $L_{ab}=L_c\equiv L$) sample for an external field of 0.2
(in units of quantum fluxes per plaquette) as a function of temperature
(which is measured in units of the Josephson energy of the junctions) for
three different sizes of the system: $L=8$, 16 and 24. For comparison, the
resistivity when the current is applied perpendicularly to the field is also
shown for the case $L=8$. It is clearly seen that the onset temperature for
the dissipation in the {\it c}-axis $T_p$ is higher than the corresponding
to the {\it ab}-plane. Fig.\ref{volperc}(b) shows the probability that the
vortex lines have percolated through the sample along the {\it ab}-plane. We
see a percolation transition around $T_p$ that becomes narrower the greater
the size of the system. This indicates that there exists a sharp percolation
transition in the thermodynamic limit. As an additional check, in Fig. \ref
{volperc}(b) (inset) the data of Fig. \ref{volperc}(b) are plotted vs a
rescaled variable $\tilde x$: $\tilde x=L_{ab}^\alpha \left[ 1/2-\left(
1-\exp \left( -\Delta /T\right) \right) ^{L_c}\right] $, where $\alpha =0.7$%
, and $\Delta =3.75$ are numerically found parameters. This scaling comes up
by using a simple model for the percolation \cite{tbp}. It strongly suggest
that a (percolative) thermodynamical phase transition is occurring in the
system.

By comparing Figs. \ref{volperc}(a) and (b) it can be seen that the
temperature $T_p$ where {\it c}-axis resistivity starts to be different from
zero is the same temperature at which the percolation probability becomes
finite. This indicates -as qualitatively discussed above- that the
percolation transition is a necessary condition for the existence of
dissipation in the {\it c}-direction.

In addition, we would like to have a more quantitative estimation of the
resistivity, based on the geometrical configurations of the vortex system.
This can be accomplished in the following way: Let us consider a sample of
size $L_c$ ($L_{ab}$) in the {\it c}({\it ab})-direction The resistivity $%
\rho $ of the sample in the{\it \ c}-direction is proportional to the number
of paths $n$ that cross the sample in the {\it ab}-plane per unit of area,
times the velocity $v$ this paths acquire under the external force, divided
by the external current density $j$: $\rho \sim nv/j$. The velocity $v$ is
given -using a viscous fluid argument- by the external force $F$ divided by
a total viscosity $\eta $, which is equal to a specific viscosity
coefficient, $\eta _0$ times the total length of the vortex path, that we
will call $l$, i.e., $v=F/\eta _0l$. The force $F$ is given in term of the
external current and the size of the system: $F\sim jL_{ab}$. We obtain $%
\rho \sim nL_{ab}/\eta _0l$. The coefficient $\eta _0$ depends on
temperature, however, on small ranges near the percolation threshold we will
take it as a constant. The determination of $n$ and $l$ is a difficult task,
because the percolation paths across the sample are not uniquely defined due
to the crossing of vortex lines (see Fig.\ref{dibujo}(a)). We will use the
following estimation: we assume that $n\times l\times L_{ab}\times L_c$ is
the volume $S$ of the percolation cluster in a sample of volume $%
L_{ab}\times L_{ab}\times L_c$. The value of $S$ can be easily evaluated
from the numerical simulation. We obtain $\rho \sim S/\eta _0l^2L_c$. It
remains to estimate the value of $l$. This length $l$ depends both on
temperature and the size of the system. As we said, a direct numerical
determination of $l$ is difficult due to indeterminacies at the crossing
points of the vortex lines. We will use the most crude estimation (see Fig.%
\ref{dibujo}): when the magnetic field $H$ is close to zero -i.e. $%
H<H_{cross}$, where $H_{cross}$ is a crossover field which is defined below-
we take $l\sim L_{ab}$. However, for $H>H_{cross}$, percolation proceeds via
the external field generated vortices and the length of a percolation path
is much larger, and can be estimated to be $l\sim L_cL_{ab}/H^{-1/2}$. In
this way we obtain the following scaling for the resistivity near the
percolation threshold
\begin{equation}
\label{uno}\rho \sim S/L_{ab}^2L_c\text{ \thinspace \thinspace \thinspace
\thinspace \thinspace \thinspace \thinspace \thinspace \thinspace \thinspace
\thinspace \thinspace \thinspace \thinspace \thinspace \thinspace \thinspace
\thinspace \thinspace for \thinspace \thinspace \thinspace }H<H_{cross},
\end{equation}
\begin{equation}
\label{dos}\rho \sim S/L_{ab}^2L_c^3\text{\thinspace \thinspace \thinspace
\thinspace \thinspace \thinspace \thinspace \thinspace \thinspace \thinspace
\thinspace \thinspace \thinspace \thinspace \thinspace \thinspace \thinspace
\thinspace \thinspace \thinspace \thinspace for\thinspace \thinspace
\thinspace \thinspace \thinspace }H>H_{cross}.
\end{equation}

This scaling is expected to be valid only close to the percolation threshold.

The crossover field $H_{cross}$ is estimated as $H_{cross}\sim 1/L_c^2$, and
corresponds to the zero-temperature lattice parameter of the vortex
structure being equal to the thickness of the sample. (The value of this
field is about 20 $G$ for a $1$ $\mu m$ thick sample). For the value $H=0.2$
used in Fig. \ref{volperc} we are in the case $H\gg H_{cross}$ for all the
values of $L_z$ considered.

In order to check the previous estimations, in Fig.\ref{tres}(a) we compare
the values of $\rho L_{ab}^2L_c^3$ and $S$ vs temperature when $L_{ab}$ is
varied between 16 and 30, for $H=0.2$. In Fig. \ref{tres}(b) $\rho
L_{ab}^2L_c^3$ and $S$ vs temperature are compared when $L_z$ is varied
between 12 and 24, for the same field $H=0.2$. The only free parameter of
the fitting is a global factor, which is the same in Figs. \ref{tres}(a) and
\ref{tres}(b). The agreement between the numerically calculated values and
the estimated ones close to the threshold is fairly good if we take into
account all the approximations made in order to obtain Eqs. \ref{uno} and
\ref{dos}. A more precise estimation of the resistivity using only the
geometrical configurations of vortex lines seems to be difficult because of
the following facts: The percolation paths across the sample are not
uniquely defined (see Fig. \ref{dibujo}(a)), and the real movement of vortex
lines under the external force will depend on the cutting energy. The
viscosity $\eta _0$ is not a constant, but a function of temperature. In
addition, the supposition of a phenomenological viscous motion of vortex
lines may not be accurate at low temperatures, when vortices creep.

The existence of two resistive transitions (in the {\it c}-axis and the {\it %
ab}-plane) has been experimentally observed in YBaCuO \cite{ejecexp2}. The
values of the two characteristic temperatures depend on the pinning, vortex
elasticity and magnetic field. In YBaCuO, as the thickness of the sample
increases the two temperatures become closer to each other. In our
simulations we find that the temperature at which the percolation transition
occurs decreases as $\sim 1/\ln (L_c)$ \cite{tbp}, as it can be deduced from
the scaling in Fig. \ref{volperc}(b) (inset).

The thermal excitations in the form of vortex lines crossing the sample
along the {\it ab}-plane destroy the phase coherence along the {\it c}-axis.
For $T>T_p$ the coherence length $\xi _c$ is of the order of the mean
distance between percolation paths, i.e., $\xi _c\sim L_c/n^{1/2}$. We
conclude that the mechanism that leads to the 2D-3D transition in high-$T_c$
materials with moderate anisotropy is the percolation of vortex line
perpendicular to the external field.

In summary, for a model high temperature superconductor we have shown by
using qualitative arguments and numerical simulations, that the onset of the
resistivity in the {\it c}-direction is related to a percolation transition
of vortex lines in the {\it ab}-plane. The results hold for $H\neq 0$ and $%
H=0$. A qualitative estimation of the resistivity near the threshold, and
its finite size scaling has been given. For the sizes of the isotropic
systems used, the percolation transition occurs at higher temperature than
the resistive transition in the {\it ab}-plane, and corresponds to a new
thermodynamic transition that should be characterized by new critical
exponents different from those obtained for the vortex glass transition when
current is applied parallel to the {\it ab}-plane. We expect these results
to be valid also for anisotropic systems, at least in the case of moderate
anisotropy, as in YBaCuO.

We acknowledge D. L\'opez and F. de la Cruz for helpful discusions and
critical reading of the manuscript. E. A. J. is supported by CONICET. C. A.
B. is partially supported by CONICET.

\begin{figure}
\caption{(a) Resistivity along the {\it c}-axis (solid circles) and along the
{\it ab}-plane (open circles), and (b)
probability of percolation across the sample, for a cubic lattice of size $L$
vs. temperature (in units of the Josephson energy of an individual junction).
Inset: percolation probability vs scaled temperature $\tilde x$
(see the text for definition)
Different symbols correspond to different sizes of the sample.}
\label{volperc}
\end{figure}

\begin{figure}
\caption{Schematic two-dimensional representation of percolation paths
(broken lines) of vortex lines (solid lines)
across the sample for the cases
$H=0$ (a) and $H>0$ (b). Note in (a) the two different percolation paths
with the same end points, and in (b) the percolation through the
externally generated vortex lines.}
\label{dibujo}
\end{figure}

\begin{figure}
\caption{Numerically calculated resistivity times $L_{ab}^2 L_c^3$
(solid symbols) and volume of the percolation cluster $S$ (open symbols)
vs temperature for different sizes of the sample
($L_{ab} \times L_{ab} \times L_c$), and
$H=0.2$.}
\label{tres}
\end{figure}

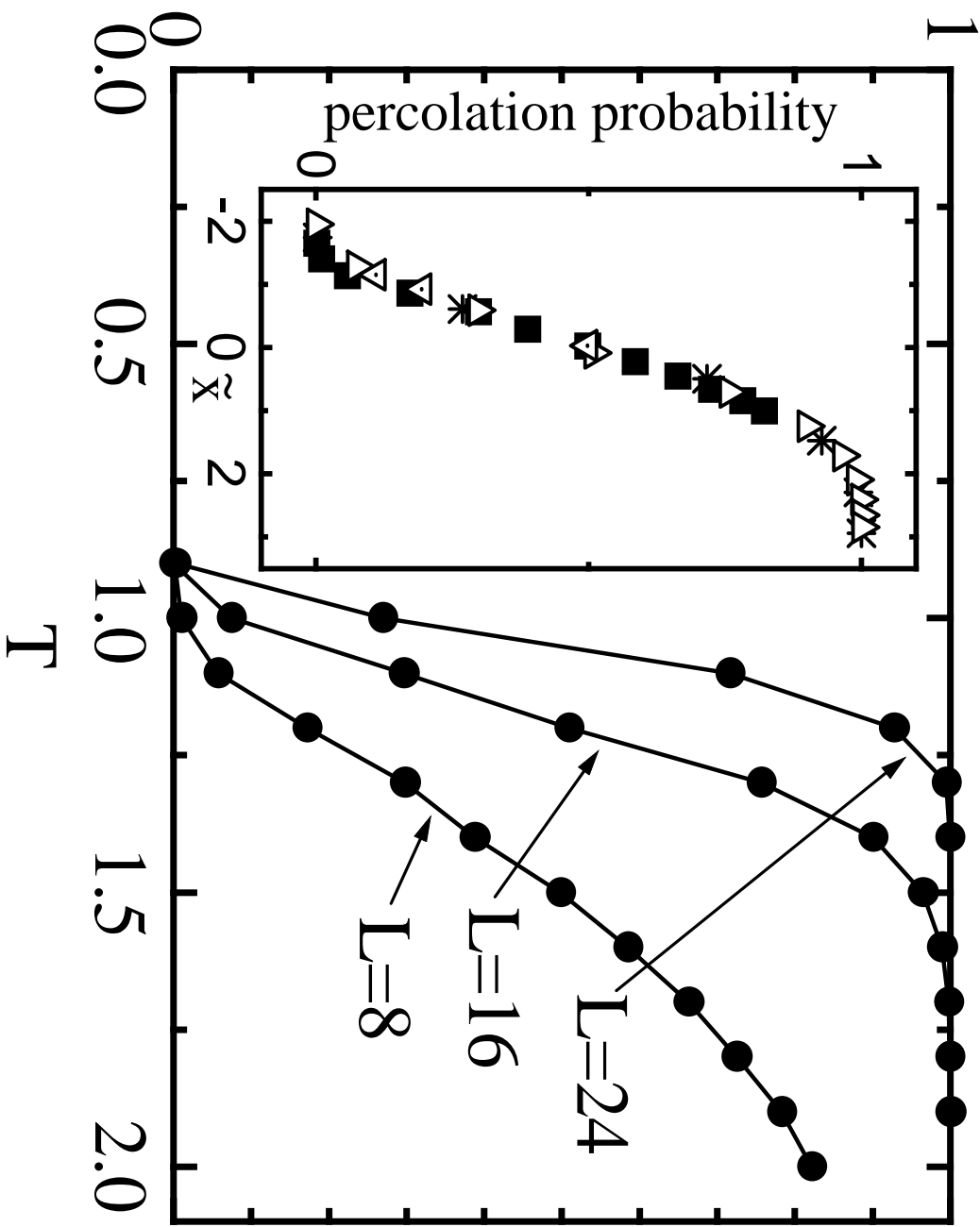
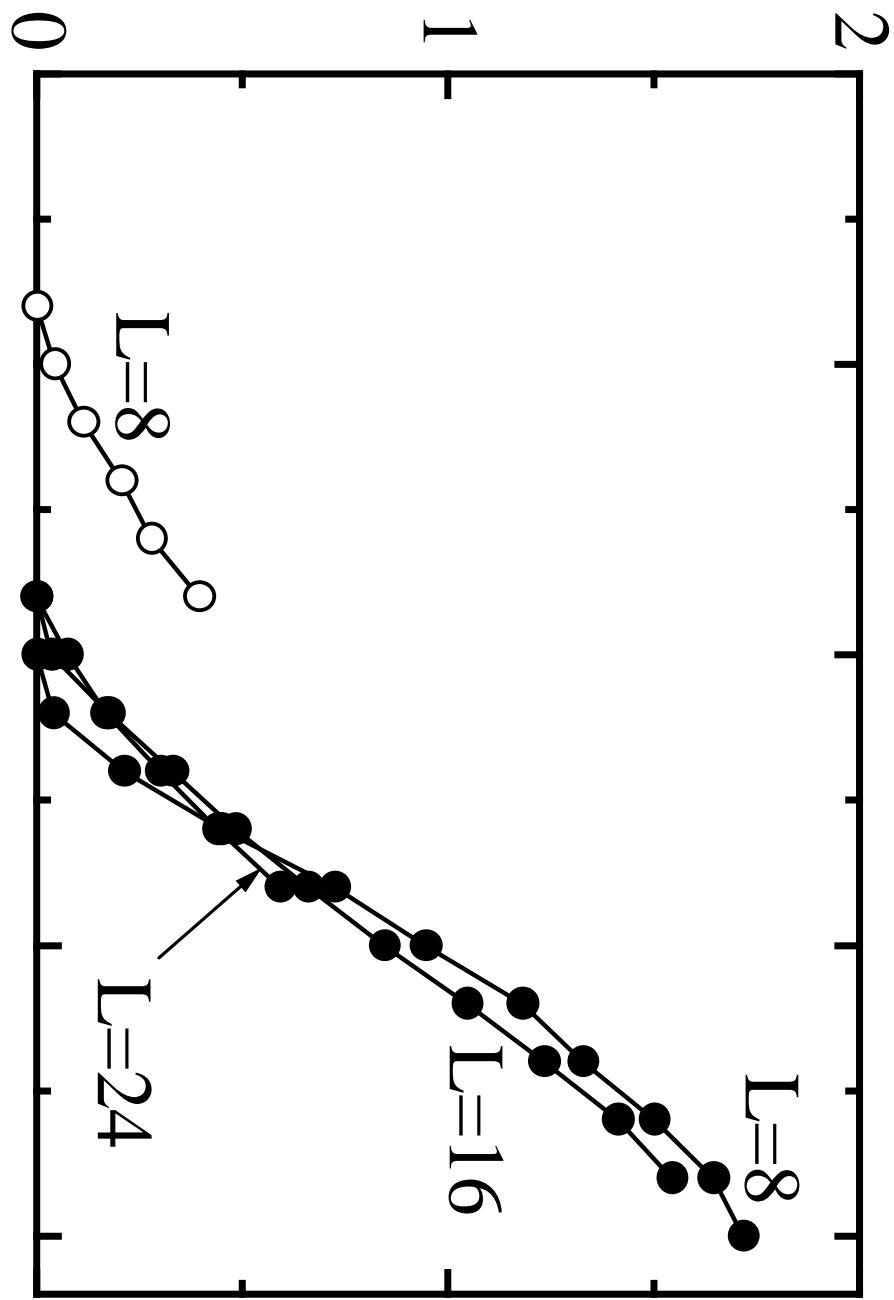
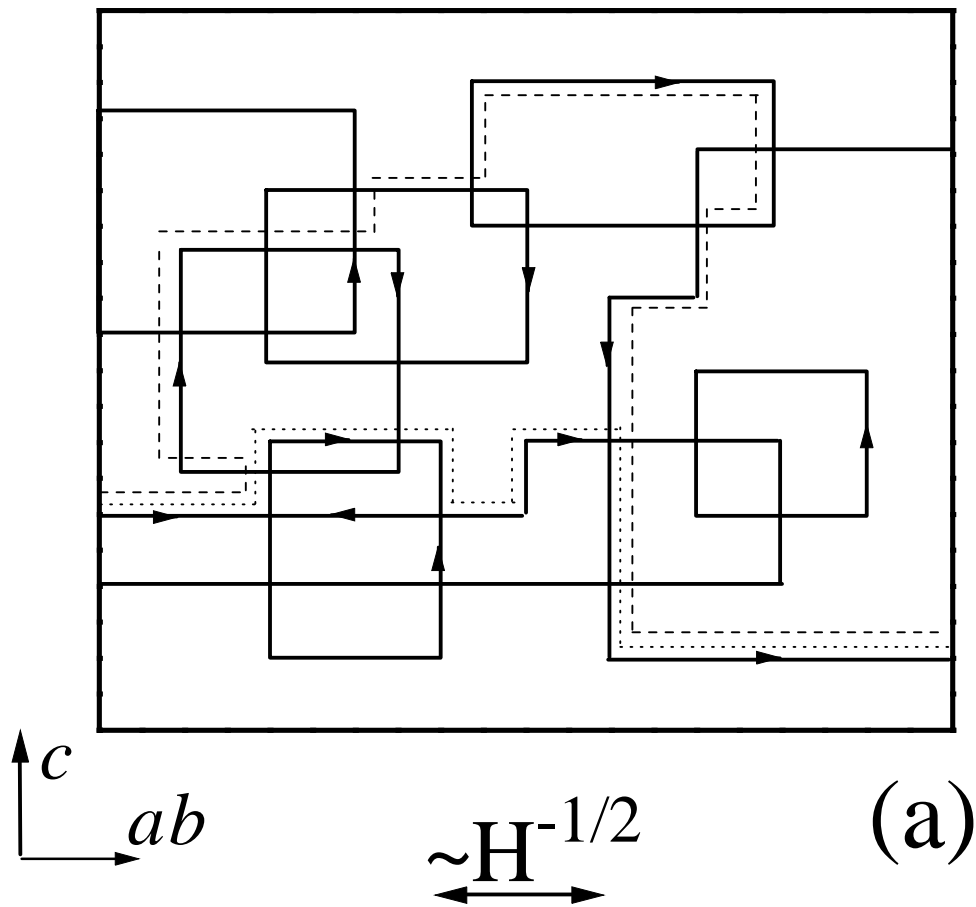
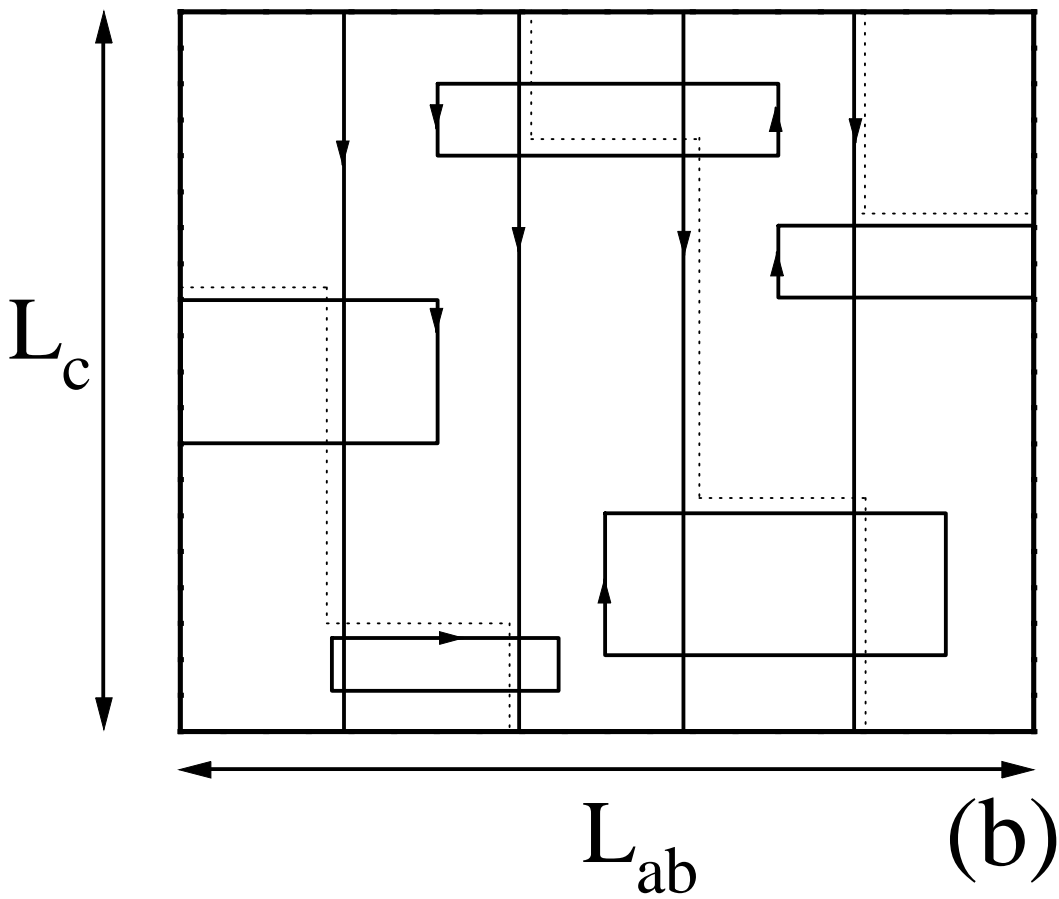
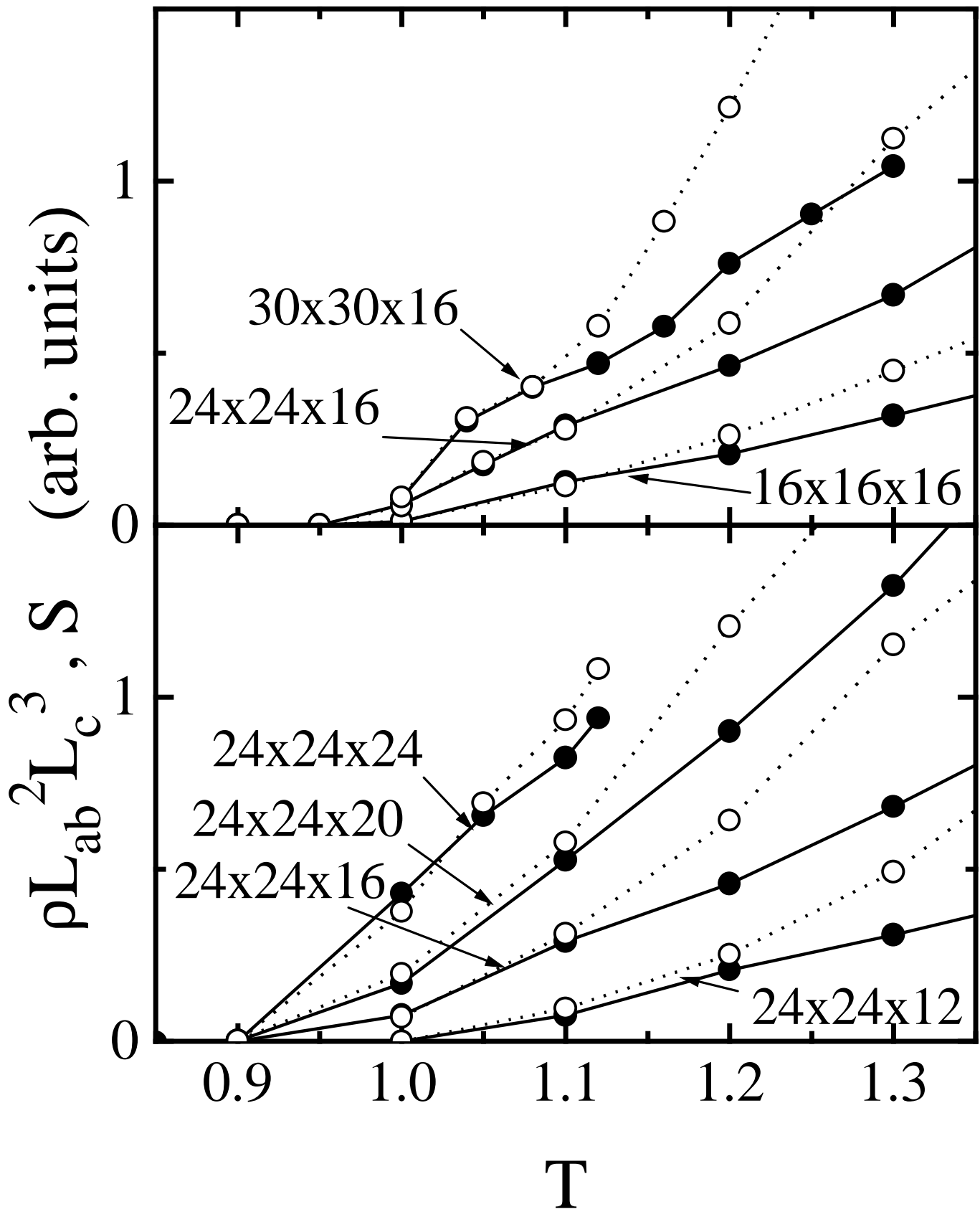
\end{document}